\documentclass[superscriptaddress,twocolumn,prx,aps,preprintnumbers,notitlepage,longbibliography,floatfix]{revtex4-1}
\usepackage[latin9]{inputenc}
\usepackage{amsmath}
\usepackage{amssymb}
\usepackage{graphicx}
\usepackage{esint}
\usepackage[scr=boondoxo, scrscaled=1.05]{mathalfa}
\allowdisplaybreaks

\makeatletter

\pdfpageheight\paperheight
\pdfpagewidth\paperwidth

\pdfoutput=1

\usepackage{epsfig}
\usepackage{graphics}

\usepackage{mciteplus}
\mciteErrorOnUnknownfalse

\makeatother

\begin{document}


\title{Gravity of gluonic fluctuations and the value of the cosmological constant}



\author{Kris Mackewicz}
\affiliation{University of Chicago}
\author{Craig Hogan}
\affiliation{University of Chicago}

\begin{abstract}
We analyze the classical linear gravitational effect  of  idealized pion-like dynamical systems, consisting of light quarks connected by attractive gluonic material with a stress-energy  $p=-\rho c^2$ in one or more dimensions.
In one orbit of a system   
      of total mass $M$,     quarks  of mass  $m<<M$ expand apart initially with $v/c\sim 1$, slow due to the gluonic attraction,  reach a maximum size  $R_0 \sim \hbar/ Mc$, then recollapse.
We solve the linearized Einstein equations and derive the effect on freely falling bodies  for two  systems: a gluonic bubble  model where  uniform gluonic stress-energy  fills a spherical volume bounded by a 2D surface   comprising the quarks' rest mass, and a  gluonic string model where a thin string connects two pointlike quarks.
      The bubble model  is shown to produce a secular mean outward residual velocity of test particles that lie within its orbit.  It is shown that  the mean gravitational repulsion of bubble-like virtual-pion vacuum fluctuations   agrees with the measured value of the cosmological constant, for a bubble with a radius equal to about twice the  pion de Broglie length. These results support the view  that the gravity of standard QCD vacuum fluctuations is the main source of cosmic acceleration.
\end{abstract}
\maketitle
\section{Introduction}

A widely repeated  calculation for the value of the cosmological constant,  based on summing the zero point fluctuations of  quantum fields, gives a famously wrong answer \cite{Weinberg:1988cp,Padmanabhan2002,Weinberg2008}:   the sum of zero-point mode-fluctuation energies up to
 a UV cutoff at mass scale $M$ leads to a cosmological constant $\Lambda$ equivalent to a mass  density of order $\rho_\Lambda\sim M^4c^3/\hbar^3$, which leads to a gravitational cosmic acceleration rate of order $H_\Lambda \sim (M/m_P)^{2}/t_P$, or a cosmological constant
 \begin{equation}\label{standard}
     \Lambda\sim (M/m_P)^{4}/t_P^2,
 \end{equation}
 where  $t_P=\sqrt{\hbar G/c^5}$ denotes the Planck time. For $M$ equal to the Planck mass $m_P= \sqrt{\hbar c/G}$, the predicted value of $\Lambda$ is larger than the observed value by about 122 orders of magnitude. Experiments\cite{adelberger2009,Kapner2007} rule out  proposed modifications of  gravity or quantum field fluctuation amplitudes with a cutoff at the milli-eV mass scale  that would give the correct value of $\Lambda$.

For this reason, it is widely agreed that there must be a basic conceptual error in the way this calculation is formulated. There needs to be a basic symmetry of quantum gravity that makes the gravitation of vacuum field fluctuations nearly vanish, and also a mechanism to account for the nonzero measured value of the actual cosmological constant.

 One possibility is that
   symmetries  of quantum geometry make $\Lambda$ exactly vanish for pointlike  particles, but allow a small nonzero   $\Lambda$ from the gravity of nonlocal vacuum fluctuation states of  interacting fields.  In this case, the value of $\rho_\Lambda$ would be much less than the Planck value quoted above, suppressed by a power of the field energy scale.
   A  long-studied example is the hypothesis   \cite{1967JETPL...6..316Z,Z68,Schutzhold:2002pr,Bjorken:2002sr,Brodsky2009,Bjorken:2010qx,Klinkhamer2009,Poplawski2010,Hogan:2020aow} is  that
  the  cosmological constant arises from quantum fluctuations in the strong interaction vacuum.

Studies of this hypothesis have generally sought to compute the expected low-energy energy momentum tensor from the system of QCD quantum fields. 
In this paper, we instead analyze the system geometrically, using classical gravitational models.
We estimate the gravitational effect of QCD field fluctuations by analyzing  simple idealized  classical systems whose energy-momentum structure  resembles that of pions,  the lowest-energy QCD  excitations. The energy-momentum of these systems  is dominated by the kinetic energy of pointlike quarks and massless gluons, and the nonlocal self-attractive interaction of the gluons. 
We then use these systems to estimate the gravitational effects of pion-like vacuum fluctuations, and show that they approximately agree with the measured cosmic acceleration.

In our simple models, fluctuations of gluonic  tension  produce   secular repulsive gravitation.  The energy-momentum tensor of a homogeneous condensate of massless gluons in localized virtual fluctuations  takes a form
proportional to the metric, with pressure and density  related by  $p= -\rho c^2$ in one or more dimensions. For more than one dimension,   this equation of state violates the strong energy condition, so its gravitational effect is repulsive.  In field language, this behavior for gluonic fluctuations in strongly-interacting QCD vacua  arises from the gravitational effect of a trace anomaly\cite{Schutzhold:2002pr}. The one-dimensional case is  also familiar from early models of pions which modeled strong interactions as strings.

Like early phenomenological models of hadrons,  our  analysis does not provide a rigorous connection to QCD field degrees of freedom.  However, it  provides a simple classical  model for gravitational effects of  nonlinear QCD fluctuations, and shows how they depend critically on nonlocal causal coherence of field states in more than one dimension.  It   provides physical insights into how the gravity of vacuum fluctuations works at a microscopic level, in particular   the reason for the small value of the cosmological constant.
 Simply put, the QCD-bubble model predicts that {\it cosmic acceleration has about the same magnitude (with opposite sign) as Newtonian gravitational acceleration at the edge of a proton.}



\section{Causally coherent gravity of  gluonic fluctuations}

\begin{figure}
  \centering
\includegraphics[width=0.4\textwidth]{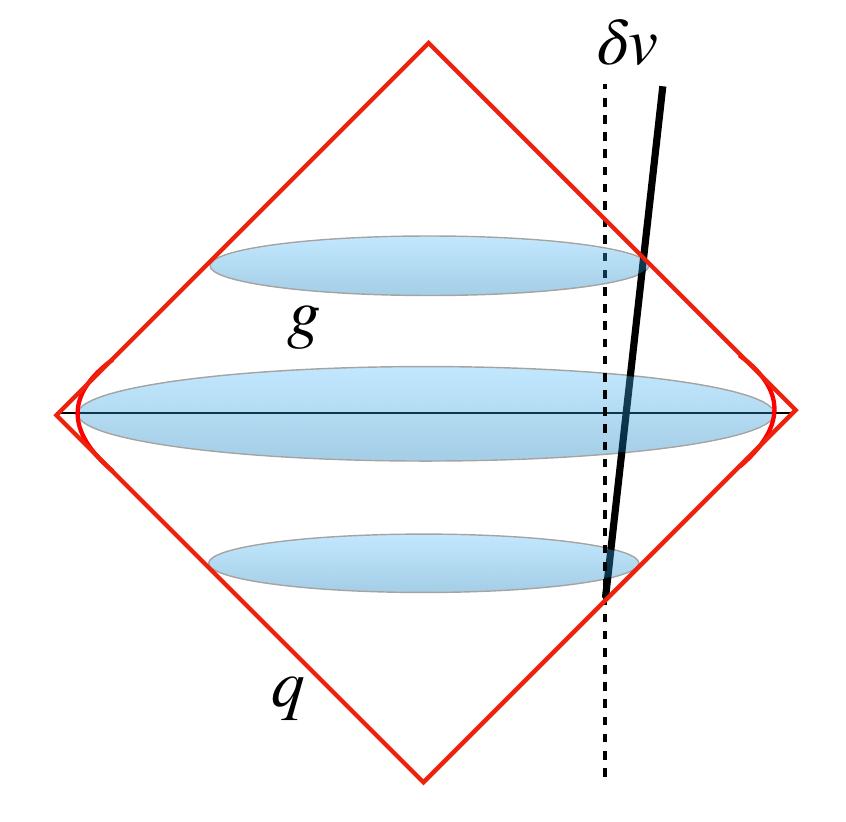}
  \caption{Spacetime diagram of  the gluonic bubble model.  The  diamond represents a  bubble  of total mass $M$ filled with gluonic material $g$ with $p=-\rho c^2$ and gravitational timescale $T_0$, with a spherical quark shell $q$ of mass $m<<M$ on a nearly-null trajectory. The shell first propagates outwards to maximum radius $R_0$, then collapses inwards, separated by a small nonrelativistic reversal region. The timelike world line  represents  a freely falling body. The gravity of the gluonic matter produces an outwards residual velocity $\delta v_g\sim + R_0^2/cT_0^2$ during the time a body spends within the bubble (Eq. \ref{gkick}). 
  \label{bubblefig}}
\end{figure}

We study the dynamics and gravity of two idealized models with different geometries.
   The first model, shown in Fig. (\ref{bubblefig}), is a bubble: a spherical volume of gluonic matter is approximated by a uniform isotropic tension and density with the Lorentz-invariant relationship 
   \begin{equation}
       p= -\rho c^2
   \end{equation}
   in  three dimensions, bounded by a uniform shell of dustlike quark material of constant total mass.
   The other model, discussed in the Appendix, is a more  traditional  idealized model  of pions, where 
a straight gluonic string with $p= -\rho c^2$ in one dimension joins two  light pointlike  quarks.
The two systems have similar dynamics: they start at small radius with a large $\gamma$  factor, expand to a maximum size determined by the masses of the quarks and the tension of the gluons, then recollapse.

As discussed below, tension in more than one dimension is required for gravity to produce cosmic acceleration, so we focus on the bubble model.
Since it is spherically symmetric,  gravity  outside the bubble  is simply a Schwarzschild metric. Inside the bubble, the  effect of the quarks on a test particle resembles displacements by a null shock on a causal diamond, whose outwards and inwards gravitational displacements cancel over a whole orbit. The main gravitational effect in the interior is from the  gluonic matter.

We find that gravity inside the  bubble  produces a mean  repulsive residual velocity, in the sense that on average it causes test particles within the orbit  to accelerate systematically apart from each other in the radial direction. 
Ultimately this unique behavior  can be traced to the exotic nature of the source, whose mass-energy, dominated for much of its orbit by the gluonic matter with $p= -\rho c^2$,  violates the strong energy condition.  
 (The string model, which does not violate this condition, also creates repulsive gravitational impulses along some directions, but not in a global average.)

We then adapt the classical model  to  estimate  the mean gravitational  effect of  QCD vacuum fluctuations.
Gravitationally repulsive virtual gluonic material is borrowed from  vacuum, so its gravitational effect  only extends over a compact causal diamond with a radius  $\sim \hbar/m_\pi c$ determined by the pion mass $m_\pi$.
 Since all of space in a sense lies ``inside a virtual bubble'', 
 this model leads to a simple picture of how cosmic repulsion works. Within  the causal diamond of a fluctuating  bubble, test bodies on one side of the bubble accelerate away from the center, and if quantum gravity is causally coherent, also away from the entire future light cone    beyond the center.  The acceleration is approximately the Newtonian gravitational acceleration for a mass with the  bubble density and bubble radius, rather than a cosmic radius. 
  As discussed further below, virtual  bubbles thus create gravitational fluctuations whose secular gravitational repulsion is much smaller than 
  Eq.(\ref{standard}):
  \begin{equation}\label{newmodel}
     \Lambda\sim (m_\pi/m_P)^{6}/t_P^2.
 \end{equation}
 
As shown below, the bubble estimate approximately agrees with the measured cosmological constant  for  parameters similar to  physical pions: for  mass $M= m_\pi$, it requires a bubble radius $R_0\sim 2.0 \ \hbar/m_\pi c$, about two femtometers. Such close agreement is remarkable, since the model is idealized in several important ways.  For example, a smaller radius  would be expected from the fact that real QCD fluctuations do not have a maximally-repulsive isotropic equation of state; their gravity would be expected to behave like something in between the bubble and the string.

The model provides a  well controlled connection, albeit still idealized,  between  the measured properties of pions and the measured cosmological constant. 
The rather close agreement, based on a simple correspondence argument and a highly idealized model system, suggests that if gravitational states of the field vacuum are coherent in causal diamonds, an absolute value for the cosmological constant can in principle  be derived from properties of Standard Model fields. 
Realistic numerical studies of gravitational effects from the QCD vacuum would not require a theory of quantum gravity, but would require a coherent nonlocal calculation of expected  mass-energy flows in the vacuum state.


\section{Gluonic bubble model}

\subsection{Bubble model with light quarks}

The idealized ``gluonic bubble'' model (Fig. \ref{bubblefig}) is a spherical ball of total mass $M$, filled with uniform gluon gas
of the unique Lorentz-invariant form $p_0=-\rho_0 c^2$,  surrounded by a thin sheet of pressureless quark dust of total mass $m$ on the surface. 
It captures the nonlocal, nonabelian self-tension of the gluon fields in an idealized way that complements the 1D string model more commonly used for pions.  It allows for a solution of the Einstein equations and derivation of gravitational effects
for an isotropic pressure in 3D.
We call it a bubble model to differentiate from the bag model, an idealized picture of a stable nucleon in a confining vacuum. The bubble model, like the string model, is an idealized picture of the dynamical mass-energy of QCD fields in a pion-like state, designed to approximate the virtual fluctuating energy flows of the QCD vacuum.

The quarklike surface of the bubble is dust, that is,  it has no tension or pressure, and is infinitesimally thin.   Its mass is constant as it expands, so the mass density thins out, and the inwards acceleration from  the constant gluonic tension increases.  The equation of motion is thus not the same as the string model, but the solutions are similar.  For light quarks $m<<M$, the bulk of the orbit is relativistic inwards or outwards motion. There is a brief turnaround near maximum expansion where the velocities are much less than $c$.

\subsection{Gravitational velocity kick from a  bubble orbit with light quarks}

In the bubble model, there is no gravitational radiation, so the outgoing and incoming parts of the orbit are identical under time reversal.
The inwards and outwards shocks from the passage of the quark surface identically cancel, so there is no residual gravitational effect of the quark surface on the motion of test particles, apart from those of a ``background'' Schwarzschild solution of mass $M$, which is the space-time outside the bubble.


However,  worldlines that pass through the interior of the bubble's causal diamond accumulate outwards acceleration while they are inside. 
The mean gravitational effect on  test bodies during the time that they pass within the volume of the bubble leaves behind an outwards ``residual velocity'' whose mean cumulative effect resembles cosmic acceleration.

It is a well known result in general relativity that the  gravitational acceleration at radius $r$   relative to the center of a homogeneous sphere is
\begin{equation} 
 a(r)  =  \dot  v  = - (4\pi/3) G r (\rho + 3p/c^2).
\end{equation}
This Newtonian  weak field limit is valid for  a system much smaller than the Schwarzchild radius of the contained mass. For empty space outside the sphere, the solution is Schwarzchild so it approaches flat space at large radii.  In the opposite limit where matter uniformly fills a large volume, the exact solutions are FRW cosmologies. 

The  effect of general relativity is captured by the last term, the Newtonian gravitational effect of pressure.  The large negative pressure within the volume of a gluon bubble leads to  a net positive acceleration or gravitational repulsion at radius $r$, 
\begin{equation}\label{acceleration}
 a_g(r)  =  \dot  v =  +  r/T_0^2,
\end{equation}
where we have defined a gravitational timescale 
\begin{equation}\label{gravtime}
    T_0\equiv (8\pi G\rho_0/3)^{-1/2},
\end{equation}
for a gluonic bubble of maximum radius $R_0$ and  density
\begin{equation}
 \rho_0= (M-m) (4\pi R_0^3/3)^{-1}.
\end{equation}

\begin{figure}
    \centering
    \includegraphics[width=\linewidth]{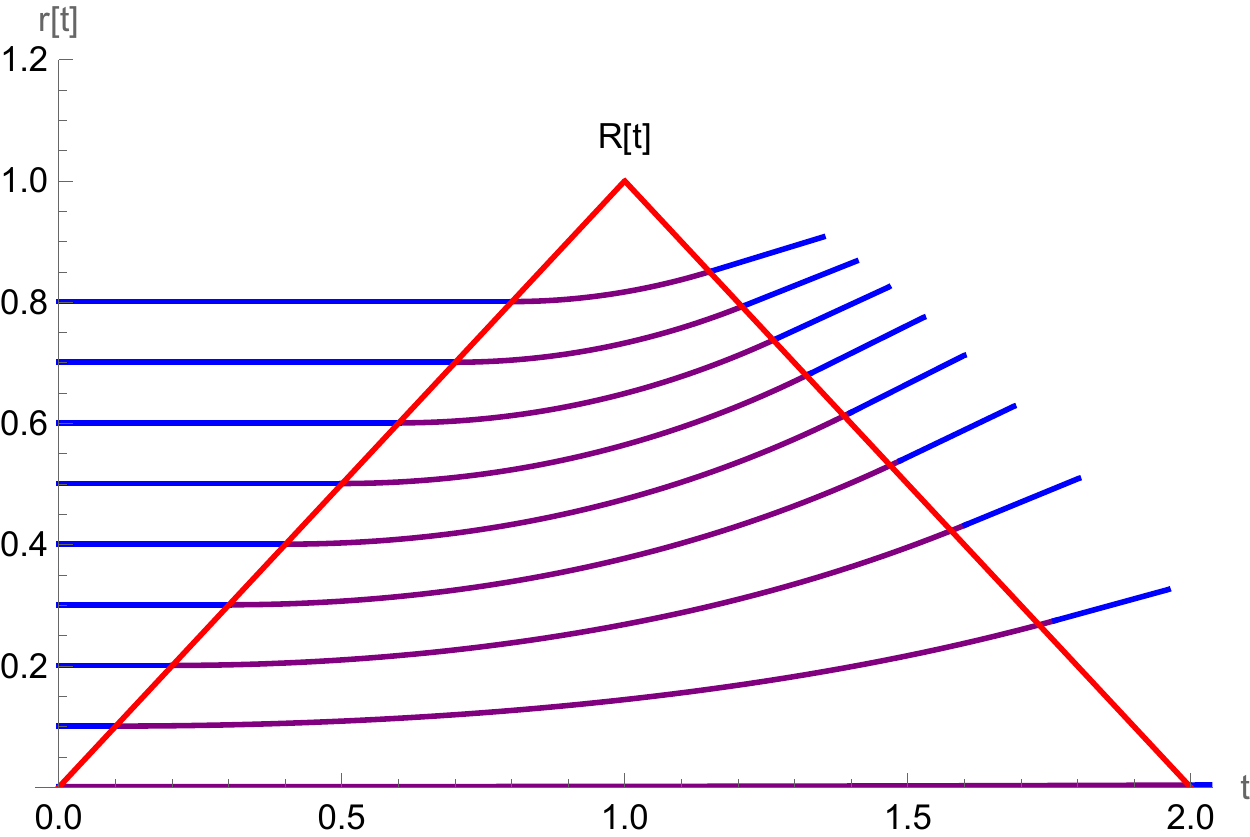}
    \caption{Trajectories of test particles (blue/purple) accelerating within the causal diamond (red) of a  bubble with $m<<M$ and $R_0=1$, according to Eq. (\ref{acceleration}). The acceleration increases with increasing $r$, but the total time inside the bubble decreases with increasing $r$, so the maximum residual velocity (Eq. \ref{vgkick}) occurs at $r=R_0/2$. For illustration, the motion of the test particles is exaggerated in this plot by a factor of $R_0^2/T_0^2$ compared to that of the bubble wall, or about 40 orders of magnitude for QCD fluctuations.}
    \label{fig:testparticletraj}
\end{figure}

In the light quark limit $m<<M$, we can   ignore the short turnaround part of the orbit.
A worldline at radius $r$ spends a time $\tau_g(r)= 2(R_0-r)/c$ inside the  bubble.
The outward velocity  accumulated at radius $r$ during this time is  
\begin{equation}\label{vgkick}
    \delta v(r)_g=\tau_g(r) a(r)_g=  + 2 r(R_0-r)/c T_0^2
\end{equation}
This quantity vanishes both  at $r=R_0$ and $r=0$, so  there is no residual velocity kick for a world line on the maximal boundary of the bubble or at the origin. 
In between, the residual velocity is positive,  with a maximum value at $r=R_0/2$,
\begin{equation}\label{gkick}
    \delta v_g(r=R_0/2)= +   R_0^2/cT_0^2.
\end{equation}
After a bubble orbit, two particles on opposite sides of the center are moving apart by the sum of their two kicks. Several sample trajectories are shown in Fig. (\ref{fig:testparticletraj}).

In the classical solution, this outwards velocity kick is combined with an inwards  acceleration accumulated  while a world line lies outside the bubble, where the metric is a Schwarzschild solution of mass $M$. The source in this regime includes the gravity of the (mostly relativistic)  quark material as well as the total mass of the gluonic material. 
As noted above, the quark contribution vanishes in the limit of light quarks, since the inwards and outwards shocks cancel.


\subsection{Gluonic bubble  solution for $m\ne 0$}

To rigorously solve for the dynamics of the bubble, we will apply the Israel junction conditions following \cite{Israel1990}. We will begin by considering gluing two general spacetimes, then restrict ourselves to the spherically symmetric case, and finally further restrict to the weak field regime (still allowing for the possibility of relativistic velocities). In this section we work in units where $c=1$ to avoid keeping track of factors of $c$ when raising and lowering tensor indices with the metric. Consider two distinct spacetime manifolds $\mathcal{M}_+,\mathcal{M}_-$ with associated metrics $g^+_{\alpha\beta}(x^{\mu}_+), g^-_{\alpha\beta}(x^{\mu}_-)$. The two spacetimes are bounded by hypersurfaces $\Sigma_+,\Sigma_-$ with induced metrics $g^+_{ab},g^-_{ab}$ ($a,b=1,2,3$). We can glue the spacetimes together by making the identification $\Sigma_+=\Sigma_-=\Sigma$ with intrinsic coordinates $\xi^a$. 

We can construct a tetrad of vectors $n^{\mu},e^{\mu}_{(a)}$ ($\mu=0,1,2,3$) (which can be defined in both spacetime regions) satisfying
\begin{equation}
    \left. n_{\mu}n^{\mu}|_+=n_{\mu}n^{\mu}\right|_-=1 \quad, \quad \left. n_{\mu}e^{\mu}_{(a)}\right|_{\pm} = 0
\end{equation}
where the vectors $e^{\mu}_{(a)}$ are adapted to the hypersurface $\Sigma$ such that
\begin{equation}
    g_{ab}^{\pm} = g_{\alpha\beta}\left.  e^{\alpha}_{(a)}e^{\beta}_{(b)}\right|_{\pm}
\end{equation}
We can parametrically define $\Sigma$ such that $\Phi\equiv R(t)-r=0$ on the hypersurface. This gives a natural identification of hypersurfaces of $\Phi>0$ with $\mathcal{M}_+$ and hypersurfaces of $\Phi<0$ with $\mathcal{M}_-$. We can then define $n^{\mu}$ to be normal to surfaces of constant $\Phi$ such that $n_{\mu}=\alpha^{-1} \partial_{\mu}\Phi$, where $\alpha$ is chosen to ensure normalization. 

One can show that by appropriate choice of intrinsic coordinates $\xi^a$ we can make $g^+_{ab}(\xi)=g^-_{ab}(\xi)=g_{ab}(\xi)$. However, there will be a discontinuous jump in the normal extrinsic curvature defined by
\begin{equation}\label{extrinsic}
    K_{ab} = -n_{\mu} e^{\nu}_{(b)}\nabla_{\nu} e^{\mu}_{(a)}
\end{equation}
In Newtonian gravity, this gives rise to the familiar jump in the normal derivative of the Newtonian potential. The induced surface stress-energy can be related to the jump in the extrinsic curvature by an analog of the Einstein equations
\begin{equation}\label{shellstressenergy}
    -8\pi \left(S_{ab}-\frac{1}{2}g_{ab}S\right) = \left[K_{ab}\right]
\end{equation}
where $[F]$ denotes $(F_+-F_-)\left.\right|_{\Sigma}$, i.e. the difference in $F$ across the hypersurface $\Sigma$. The full stress-energy tensor restricted to the hypersurface $\Sigma$ is then given by
\begin{equation}
    T^{\mu\nu}_{\Sigma} = -S^{ab}e^{\mu}_{(a)}e^{\nu}_{(b)}|\alpha| \delta (\Phi)
\end{equation}

Now, we will restrict our attention to spherically symmetric spacetimes. In Eddington-Finkelstein coordinates, the metric can be written as
\begin{equation}\label{sphericalsymmetric}
    ds^2 = e^{\psi} du (f e^{\psi} du + 2 \zeta dr) + r^2 d\Omega
\end{equation}
where $u=t-\zeta r^*, d r^*/d r=1/f$ and
$f(u,r)=1-2m(u,r)/r$. $\zeta=\pm 1$ denotes whether the hypersurface $\Sigma$ is moving outward (increasing $r$) or inward. The Einstein equations then give us differential equations for the functions $m,\psi$. 
\begin{align}
    \partial_u m &= 4\pi r^2 T^r_{\,u} \label{partialum} \\
    \partial_r m &= -4\pi r^2 T^u_{\, u} \label{partialrm}\\
    \partial_r \psi &= 4\pi r T_{rr} \label{partialrpsi}
\end{align}
For the bubble model being considered, the stress-energy inside of the shell is that of a de-Sitter spacetime with positive cosmological constant and stress-energy proportional to the spacetime metric.
\begin{equation} \label{falsevacenergy}
    T^{dS}_{\mu\nu}= - \rho g_{\mu \nu} 
\end{equation}
In the Eddington-Finkelstein coordinates we find
\begin{equation}
    T^u_{\, u} = T^r_{\, r} = T^{\theta}_{\, \theta}=T^{\phi}_{\,\phi} = -\rho 
\end{equation}
Therefore we get that $T_{rr}=0,T^r_{\, u}=0$, which by eq. (\ref{partialrpsi}),(\ref{partialum}) imply $ \psi=0,f=f(r)$. Solving eq. (\ref{partialrm}) assuming the exterior region to be Schwarzschild, we find
\begin{align}
    f_+ = 1-\frac{8\pi}{3}\rho r^2  \\
    f_- = 1- \frac{2M}{r}
\end{align}
Now we may explicitly define the tetrad in terms of the chosen coordinates. We will switch back to using the more familiar $t,r$ coordinates. 
\begin{align}
    n_{\mu} = -\frac{1}{(f-f^{-1} \dot{R}^2)^{1/2}} \left(\delta^r_{\mu}+\dot{R}\delta^t_{\mu}\right)
\end{align}
where the over-dot signifies the derivative with respect to coordinate time $t$. The intrinsic metric for the timelike spherical shell is given by
\begin{equation}
    ds^2_{\Sigma} = -d\lambda^2 + r^2d\Omega^2
\end{equation}
where $\lambda$ is the proper time of a co-moving observer on the shell. Then we can choose the rest of our tetrad vectors to be
\begin{align}
    e^{\mu}_{(1)} &= \frac{1}{(f-f^{-1} \dot{R}^2)^{1/2}}\left(\delta^{\mu}_t+\dot{R}\delta^{\mu_r}\right) \\
    e^{\mu}_{(2)} &= \delta^{\mu}_{\theta} \\
    e^{\mu}_{(3)} &= \delta^{\mu}_{\phi}
\end{align}
One can compute the extrinsic curvature using eq. (\ref{extrinsic}), which produces the following stress-energy on the spherical shell:
\begin{equation}
    -S^{ab} = \sigma \delta^a_1 \delta^b_1
\end{equation}
\begin{equation}\label{shellenergy}
    T^{\mu\nu}_{\Sigma} = |\alpha| \sigma e^{\mu}_{(1)}e^{\nu}_{(1)}\delta(r-R(t)) 
\end{equation}
where $\sigma$ is the surface energy density of the shell given by
\begin{equation}
    \sigma = -\zeta \frac{[m]}{4\pi r^2}
\end{equation}
The form of eq. (\ref{shellenergy}) is that of a pressureless dust in the rest frame of the shell. In the weak-field limit this reduces to the usual form
\begin{equation} \label{weakfieldenergy}
    T^{\mu\nu}_{\Sigma} \approx \frac{\gamma m}{4\pi r^2} v^{\mu}v^{\nu} \delta(r-R(t))
\end{equation}
where $v^{\mu}=dx^{\mu}/dt, \gamma=(1-\dot{R}^2)^{-1/2}$. The evolution of the shell radius $R(t)$ is determined by conservation of stress-energy and is given by
\begin{equation}
    \left[{\rm sgn}(n^{\nu}\partial_{\mu}r)(f+(dR/d\lambda)^2)^{1/2}\right] = -\frac{M}{r}
\end{equation}
Using the definition for $n^{\mu}$ and $f_{\pm}$ we find
\begin{equation}
    \Gamma m = M - 
    \frac{4}{3}\pi r^3\rho
\end{equation}
where $\Gamma$ is defined by
\begin{equation}
    2\Gamma = (f_++(dR/d\lambda)^2)^{1/2}+(f_-+(dR/d\lambda)^2)^{1/2}
\end{equation}
In the weak field limit with $dR/d\lambda>>1$ this reduces to
\begin{equation}
    \Gamma \approx \gamma = (1-\dot{R}^2)^{-1/2}
\end{equation}
which is consistent with the expected mass/energy conservation law. 

Next, let us define the glued metric over the entire spacetime by
\begin{equation}
    \tilde{g}_{\mu\nu} = g^+_{\mu\nu}\Theta(\Phi)+g^-_{\mu\nu}\Theta(-\Phi)
\end{equation}
where $\Theta(x)$ is the heaviside step function, and $\Phi$ again parameterizes the hypersurface $\Sigma$. Other quantities with an over tilde are defined to have a similar meaning. Since the metric can be made continuous along $\Sigma$, we get

\begin{align} 
    \partial_{\alpha} \tilde{g}_{\mu\nu} &= \partial_{\alpha}g^+_{\mu\nu}\theta(\Phi)+\partial_{\alpha}g^-_{\mu\nu}\theta(-\Phi)+[g_{\mu\nu}]\delta(\Phi)\partial_{\alpha}\Phi \nonumber \\
    &= \widetilde{\partial_{\alpha} g_{\mu\nu}} \label{derivjump}
\end{align}
A direct consequence of eq. (\ref{derivjump}) is that the Christoffel symbols suffer a step discontinuity, but there is no $\delta(\Phi)$ contribution. Therefore, the radial acceleration experience by an observer crossing the shell does not produce an instantaneous displacement kick.

However, the Riemann tensor does contain such a delta function contribution, indicating that two nearby test bodies would experience an instantaneous relative velocity kick.
\begin{equation}
    R^{\alpha}_{\,\,\beta\mu\nu} =  \tilde{R}^{\alpha}_{\,\,\beta\mu\nu} - 2[\Gamma^{\alpha}_{\beta [\mu}]n_{\nu]}\alpha \delta(\Phi)
\end{equation}

From this point forward we shall operate in the weak-field limit exclusively and assume $\gamma>>1,m<<M$. The full stress-energy tensor for a gluonic bubble connected to a shell of mass $m>0$ is given by eq. (\ref{falsevacenergy}) and (\ref{weakfieldenergy})
\begin{equation}\label{bubblestressenergy}
    T_{\mu\nu} = -\rho g_{\mu\nu}\Theta(R(t)-r)+\frac{\gamma m}{4\pi R(t)^2}\delta(r-R(t)) v_{\mu}v_{\nu}
\end{equation}
where $\rho$ is the energy density of the gluonic region which acts as a perfect fluid with $p=-\rho$ and $v_{\mu}=(-1,\dot{R},0,0)$ is the four velocity of the shell. Conservation of stress-energy gives the equation of motion for the surface of the bubble:
\begin{equation}\label{bubbleEOM}
    \ddot{R} = - \frac{4\pi \rho R(t)^2}{m \gamma^3}
\end{equation}
This still assumes the bubble is small so the gravitational effect on the wall is negligible compared to the gluon tension.

\begin{figure}
    \centering
    \includegraphics[width=0.45\textwidth]{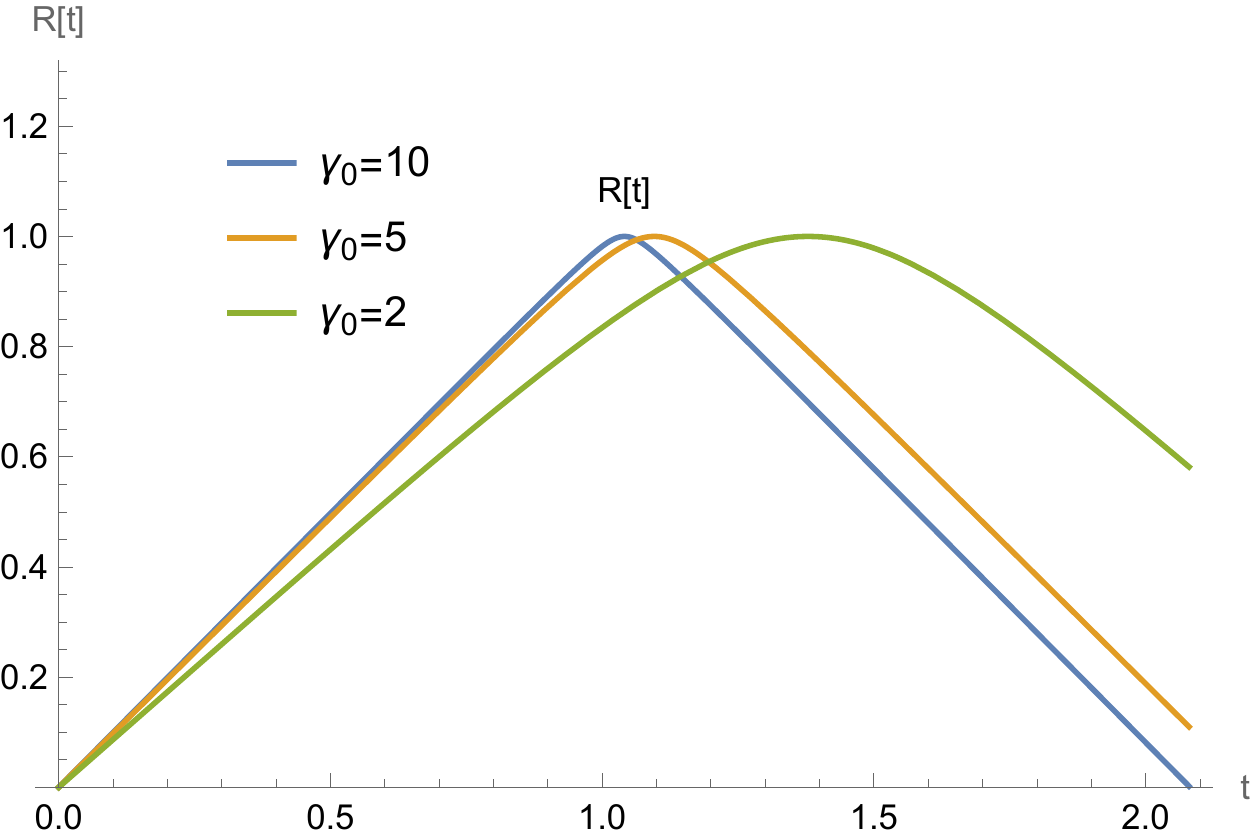}
    \caption{Trajectory of the shell $R(t)$  (solution to eq. (\ref{bubbleEOM})) for fixed bubble tension and variable initial boost factor $\gamma_0$. For $\gamma_0\sim 1$, the trajectory is no longer predominantly nearly null, and the turnaround is less abrupt. The units of the spatial axis have been re-scaled by $\hbar/M c$ while the units of the time axis have been re-scaled by $\hbar/M c^2$.}
    \label{fig:bubbleradiusrho}
\end{figure}

\begin{figure}
    \centering
    \includegraphics[width=0.45\textwidth]{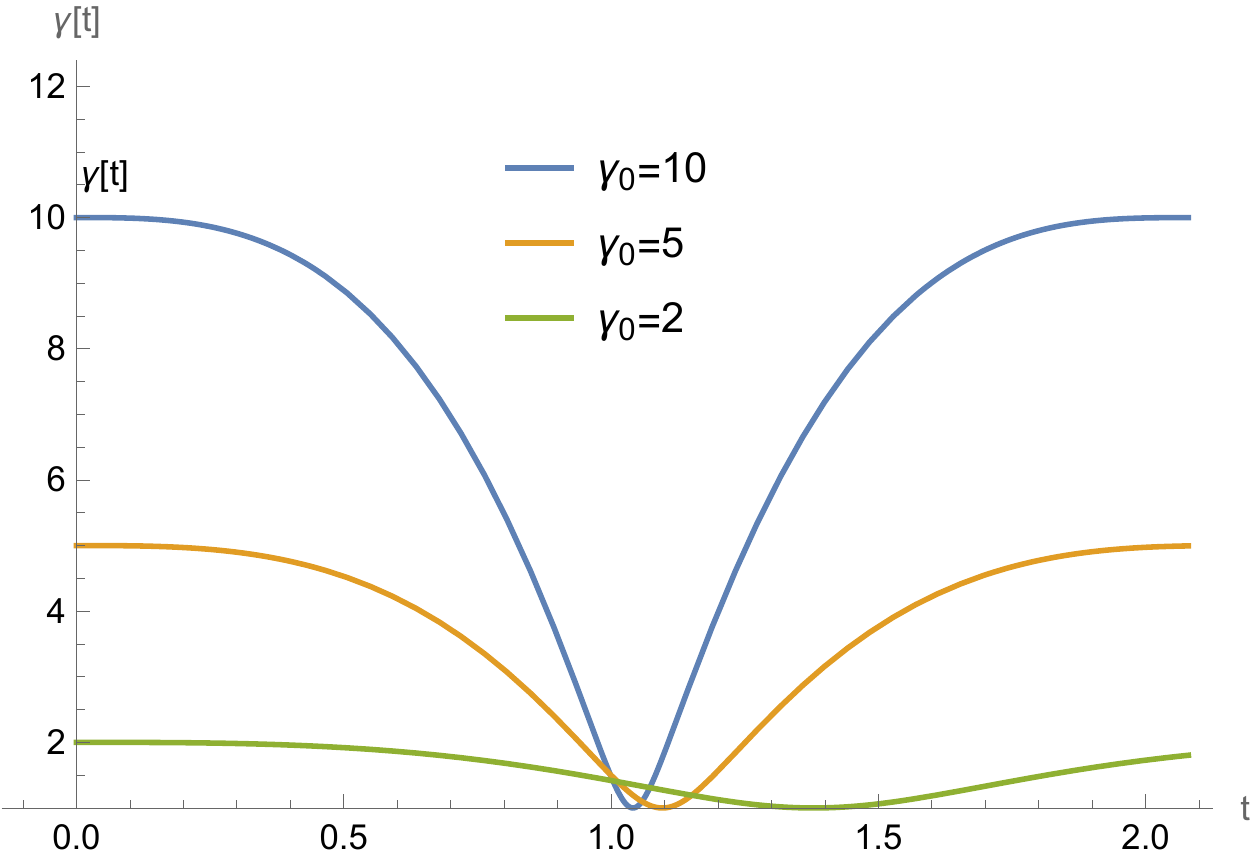}
    \caption{$\gamma$ boost factor  $\gamma=(1-\dot{R}^2)^{-1/2}$ of the shell for fixed bubble tension and variable initial boost factor $\gamma_0$. The early and late time behavior are approximated in eq. (\ref{gammasol}) and (\ref{subrelgammasol}).}
    \label{fig:boostfactorrho}
\end{figure}

Finding a closed-form analytic solution to this differential equation is quite difficult. It is easier to find an approximate solution for $\gamma(t)$ during the initial era and the turn-around. The total mass of the system is conserved and is given by
\begin{equation}\label{massconservation}
    M = \frac{4}{3}\pi \rho R(t)^3 + \gamma m
\end{equation}
For highly relativistic initial velocity, we have $\dot{R}\approx 1, R(t)\approx t$. During this portion of the evolution, we have
\begin{equation}\label{gammasol}
    \gamma(t)\approx \gamma_0-\frac{4}{3m}\pi \rho t^3
\end{equation}
Near the turn-around point, the shell will become sub-relativistic. The equation of motion then approximates to 
\begin{equation}\label{subrelEOM}
    \ddot{R} \approx - \frac{4\pi \rho R_0^2}{m}
\end{equation}
where $R_0$ is the maximum value of $R$ at the turn-around. The solution is
\begin{equation}\label{subrelradiussol}
    R(t) \approx R_0 - \frac{2\pi \rho R_0^2}{m} (t-t_0)^2
\end{equation}
\begin{equation}\label{subrelgammasol}
    \gamma(t) \approx  1+\frac{8\pi^2 \rho^2R_0^4}{m^2}(t-t_0)^2
\end{equation}
where $R(t_0)=R_0$. Before the turnaround, both $R(t),\gamma(t)$ are monotonic functions of time, so we can smoothly connect the cubic and quadratic regions for $\gamma(t)$. Numerical solutions to eq. (\ref{bubbleEOM}) are plotted in Fig. \ref{fig:bubbleradiusrho}. The corresponding boost factor $
\gamma$ is plotted in Fig. \ref{fig:boostfactorrho}.

In the limit of zero quark mass, the period of a single orbit of the bubble is given by $T=2R_0$. For a test body which begins at rest at position $r$, the time spent inside of the bubble interior in the zero quark mass limit is given by $\tau=2(R_0-r)$. For the case $0\neq m<<M$, the period is given by
\begin{equation}\label{bubbleperiod}
    T = 2 \int_0^{R_0} \frac{dR}{|\dot{R}|}
\end{equation}
where $\frac{4}{3}\pi \rho R_0^3 = M-m$. From eq. (\ref{massconservation}) we find that
\begin{equation}
    \dot{R} = \pm \left(1-\frac{m^2}{(M-(4/3)\pi \rho R^3)^2}\right)^{1/2}
\end{equation}
Plugging this into eq. (\ref{bubbleperiod}), making a variable substitution, and substituting $\gamma_0=M/m$ we get
\begin{equation}
    T = 2R_0 \int_0^1 \left(1-\frac{1}{\gamma_0^2(1-(1-\gamma_0^{-1})x^3)^2}\right)^{-1/2} dx
\end{equation}
Taylor expanding for $\gamma_0>>1$ the orbit period is approximately
\begin{equation}
    T\approx 2 R_0\left(1+\frac{1}{3\gamma_0}+\frac{1}{27\gamma_0^2}(\sqrt{3}\pi+9 {\rm ln}3+6{\rm ln}\gamma_0)\right)
\end{equation}
The time spent inside of the bubble by a test body to leading order in $m/M$ is given by
\begin{equation}
    \tau \approx 2(R_0-r)\left(1+\frac{m}{3M}\right)
\end{equation}
Meanwhile, the radial acceleration experienced by a test body scales as
\begin{equation}
    \dot{v} = \frac{2 M}{R_0^3}\left(1-\frac{m}{M}\right)
\end{equation}
Therefore the accumulated residual velocity after a single orbit is given by
\begin{equation}\label{mcorrection}
    \delta v \approx \frac{4M}{R_0^3}r(R_0-r)\left(1-\frac{2m}{3M}\right)
\end{equation}
Since the true quark masses would yield $\gamma_0\sim 10$, there will be a small but finite correction to the estimate for the radius of the bubble needed to produce the observed cosmic acceleration computed in section IV B. 

 \section{Cosmic acceleration}
 
 \subsection{Gravity  of virtual  fluctuations}

In our model of the gravitational effect of the vacuum, we will assume that  virtual point particles have no gravitational mass.    The whole gravitational effect of  virtual QCD fluctuations lies within bubbles, and is dominated by gluons. 
Since a virtual fluctuation has zero mean energy and ``borrows'' energy only causally,  the only gravitational  effect is internal to the causal diamond occupied by the  gluonic field fluctuation\cite{Brodsky2009}.
Thus,  nonlocal  gluonic fluctuations produce a residual velocity between particles approximated by the classical bubble solution,   given by    Eq. (\ref{gkick}) for each fluctuation on scale $R_0$.

Consider  gravitational repulsion from  a space-filling vacuum of virtual  pion-like  bubbles as a model for how the physical cosmological constant is produced by  QCD vacuum fluctuations.
  A small secular acceleration comes from  the accumulation of small mutually repulsive velocity  kicks within each orbit. The characteristic acceleration time  is
\begin{equation}\label{Rkicks}
T_\Lambda=  R_0 /\delta v_g\sim  cT_0^2/R_0,
\end{equation}
that is, it is larger than the  gravitational timescale 
$T_0$ by a factor $cT_0/R_0$.
 In Planck units,  $T_0^2\sim M^{-4}$ and $R_0\sim M^{-1}$, 
which leads to a ``cosmic'' acceleration rate
\begin{equation}\label{HM3}
 T_\Lambda^{-1}\sim M^3.
\end{equation}
This approximately agrees with the observed value of $\Lambda$. As explained in more  detail  below,
it is much smaller than  the value that would correspond to a universe filled with gluonic plasma of density $\rho_0$, or with thermal or quantum field excitations on the same scale, which is the standard estimate as in Eq. (\ref{standard}),
\begin{equation}
     T_0^{-1}\sim M^2.
\end{equation}
The  difference between the residual effect of a fluctuation, and the residual effect of a volume uniformly filled with the same material, arises because the gravitational acceleration from fluctuations comes only from the mass of material on the bubble scale $R_0$, instead of a  volume  with a gravitational radius $\sim cT_0$. 

 In this microscopic physical picture of how cosmic acceleration works, test particle trajectories, which are  shown in Fig. (\ref{fig:testparticletraj}) in flat-space coordinates, correspond to geodesics of the emergent, slightly curved cosmological metric.
 
 \subsection{Cosmic acceleration from virtual bubbles }

The bubble model allows a more precise comparison of   virtual bubble parameters with measured cosmic acceleration. Fits to  cosmological data\cite{Tanabashi:2018oca,Prat2021}   yield an estimated value $\Lambda_0$ that corresponds to   acceleration on a cosmic scale with a rate
\begin{equation}\label{rate}
 T_\Lambda^{-1} \equiv \sqrt{\frac{\Lambda}{3}} = 1.0 \times 10^{-61} \ t_P^{-1}\  \sqrt{\frac{\Lambda}{\Lambda_0}}.
\end{equation} 
(A pure-vacuum cosmology would have a Hubble radius and event horizon radius
$c/H_\Lambda=cT_\Lambda.$)
We now evaluate the bubble parameters for which  this mean cosmic  acceleration  matches the mean repulsive acceleration of test particles in  the bubble model. 

For virtual fluctuation states coherent on causal diamonds, the physical picture is that a test particle inside  any virtual bubble accelerates away from the entire universe on the opposite side of the bubble's center at the mean rate given by the bubble model. 
The mean  acceleration is  given by the mean repulsive
 velocity impulse over a bubble orbit (Eq. \ref{vgkick}), divided by the duration of the orbit $2R_0/c$, which we equate with cosmic acceleration:
 \begin{equation}\label{dotvH}
  T_\Lambda^{-1}= (1/2)  \langle\delta v_g / R_0  \rangle_B,
\end{equation}
where $\langle  \rangle_B$ denotes a volume average over the world lines that pass through the bubble. 
Since the impulse accounts for a whole orbit, an average that gives equal weight to each element of the  bubble 3-volume also accounts for the time average of the fluctuating acceleration:
\begin{equation}\label{volumeaverage}
    \langle\delta v_g (r) \rangle_B=
    \frac{\int_0^{R_0}dr r^2 \delta v_g(r)}{\int_0^{R_0}dr r^2}
\end{equation}
This weighting yields
\begin{equation}
    T_\Lambda^{-1}= 3 R_0/20 c T_0^2,
\end{equation} 
so writing the result in Planck units,
\begin{equation}
    T_0^{-2}\equiv 8\pi G\rho_0/3= 2 M^4 (R_0M)^{-3},
\end{equation}
we obtain
\begin{equation}\label{Hbubble}
    T_\Lambda^{-1}= (3/10) M^3 (R_0M)^{-2} . 
\end{equation}
Apart from the numerical coefficient, with $R_0M\sim 1$ this is the same result as the simple estimate in Eq. (\ref{HM3}).


  Combining these results, the predicted cosmological constant from virtual gluonic bubbles in Planck units is:
\begin{equation}
    \Lambda_{bubble}= 3 \mathcal{H}^2 M^6,
\end{equation}
where
\begin{equation}
  {\mathcal{H}}\equiv   (3/10) (R_0M)^{-2}.
\end{equation}
For bubbles with  the physical pion mass ($M=m_{\pi_0}= 135 {\rm MeV}$), we find
\begin{equation}\label{precisecomparison}
   \frac{\Lambda_{bubble} (M=m_\pi)}{\Lambda_0}
=\left(\frac{R_0 m_\pi c}{2.0 \ \hbar }\right)^{-4} .
\end{equation}
That is, 
for pion-mass fluctuations  to give  the right cosmological constant, 
the  one  parameter in this simple model--- the size of a bubble in units of the de Broglie wavelength for its mass--- needs to be
\begin{equation}\label{preciseradius}
    R_0 =  2.0 \ \hbar/m_\pi c,
\end{equation}
for
their mean gravity  to produce the observed  cosmic acceleration.
This idealized model shows quantitatively  how a cosmological constant close to the observed value  results  from the gravitational effect of vacuum fluctuations in QCD fields, as long as the quantum states of the fields and their gravity are coherent on causal diamonds.

\subsection{Improvements on the idealized bubble model}

The  comparison of cosmological and microscopic measurements in Eqs. (\ref{precisecomparison}),(\ref{preciseradius})  is precise, but it is not accurate:   it   is based on a highly idealized model system  
and is not expected to produce exact agreement with the physical cosmological constant.
  In the real QCD vacuum, coherent gluonic wave states have a more complex 4D structure than the bubble model.
  The  quantum wave function of virtual gluonic matter is not a homogeneous sphere, that of quarks is not a thin shell,
actual pion states are not radially homogeneous,  and virtual stress is not isotropic as in the bubble.
A more stringlike gluon state, which has less repulsive gravity, would require a smaller value of $R_0$ to match the observed cosmic acceleration.
  The estimate just given also does not allow for finite quark mass, $m\ne 0$, but since
physical quark masses have $m<<m_\pi$, this difference produces only a small fractional change, as shown in the solution above (Eq. \ref{mcorrection}).

There are also ambiguities in our idealized application of the correspondence principle to virtual orbits, which depend on how quantum gravity actually works in detail.  For example,  the volume average taken  above (Eq. \ref{volumeaverage}) uses the mean acceleration of bodies relative to the center of the bubble over an orbit, but it might be more accurate to include  a directional projection of the component of radial acceleration onto the opposite hemisphere of the causal diamond. Such a projection factor would change the answer by a small numerical factor.
In principle,  nonlocally-coherent gravitational effects of vacuum QCD fluctuations could be   better approximated with an explicit calculation of nonlinear quantum field dynamics.

 
 

\subsection{Remarks}

\subsubsection{Why  QCD?}
 
 It is natural to ask, why QCD? What's special about its vacuum, compared to the other fields, that make it source the cosmological constant?
 
 In a coherent relational model of locality, there are straightforward physical reasons why the gravitational effect of  vacuum fluctuations for most standard model fields should vanish. The classical gravity of a null particle with momentum $p$ is simple: it creates a null shock with a displacement $\delta \tau = Gp/c^4$, with an
 observable portion that  depends on the location of the particle relative to observer\cite{Mackewicz2022}.
A zero-point field vacuum excitation, generated by a creation operator on an infinite plane wave mode, creates  a completely delocalized state, so according to the correspondence principle, there is no observable gravitational effect: essentially, everything ``moves together''.

This argument applies to  the Standard Model fields whose interactions and correlations fall off in the infrared.
As noted above, the vacuum fluctuations of gluons are uniquely different from those of other forces. The ``IR slavery'' of the QCD vacuum confines baryons  into bags, and leads to a finite range at the Fermi scale for strong Yukawa interactions  mediated by pions.  The same effect makes QCD vacuum gravity different from that of the other forces: the vacuum fluctuations of QCD correspond to coherent localized bubbles of energy flow  on the Fermi scale, so the argument just given for delocalized  vacuum states does not apply. 
 Renormalization required for quantum field theory fails to account correctly for gravitational entanglement of causal structure with long wavelength modes\cite{HollandsWald2004,Stamp_2015}, so it is plausible that the IR slavery of vacuum QCD fluctuations leads to  different gravitational effects from other fields.


 Outside of hadrons, the  QCD vacuum  at low temperature is a  coherent condensate, whose mean  gravitating density is negligible\cite{Brodsky2009}. 
Its fluctuations  resemble the lightest resonant excitations,  pions, which are  spatially extended  but localized.
  Nearly all of their virtual energy  comes from the massless gluon field. In the gluonic bubble model, fluctuations in the tensile gluon interaction energy produce a small but cosmologically detectable repulsive gravity.
  Estimates of the effective equation of state from field theory\cite{Schutzhold:2002pr,Bjorken:2002sr,Bjorken:2010qx,Klinkhamer2009,Poplawski2010} reproduce the estimate from our bubble model.
  
The other non-Abelian forces of the Standard Model, the weak interactions, are mediated by massive particles with a short range, and the nonlocal space-time correlations of their vacuum fluctuations are qualitatively different from QCD.  For these, the ``zero momentum mode'' of fluctuations takes the form of  a globally spatially uniform scalar condensate with homogeneous fluctuations around the minimum of an effective potential. This Higgs condensate, whose order parameter describes the low-temperature vacuum expectation value of the effective potential, apparently has zero gravitation\cite{Weinberg:1988cp}. As explained in the Appendix, the difference in gravitational effect from the space-filling QCD vacuum can be understood from an exponential suppression  of the trace anomaly at weaker coupling strength\cite{Schutzhold:2002pr}.

\subsubsection{Causal coherence of virtual  fluctuations}

The bubble model illustrates classically how
a  fluctuation could have a durable macroscopic physical effect  if    positional relationships among world lines are determined by coherent causal diamonds.
In such an emergent  relational holographic picture, 
classical locality  emerges as a consistent approximation on large scales,  based on  relationships of a causal diamond with those it is nested in.  Exact relational positions within causal diamonds are indeterminate.

In our model, systematic secular effects of  fluctuations  are assumed to lead  to a durable  effect on the  classical metric.
This hypothesis leads to the assumption used in our estimate of mean acceleration of test particles, relative to the center of the bubble. On average, the acceleration applies to test particles in relation to the future light cones on  the opposite side of a pion-like causal diamond vacuum fluctuation.
The coherence propagates local coherent acceleration to the future light cone of a microscopic causal diamond, which  leads to coherent acceleration of the same magnitude on a cosmic scale.


The bubble model illustrates concretely how  causal coherence of virtual fluctuation states is connected with the small nonzero value of the cosmological constant.
According to this scenario, the wildly wrong estimate of vacuum fluctuation density in Eq. (\ref{standard}) results from an incorrect physical interpretion of vacuum energy that does not take directional causal coherence of virtual states into account;  it arises from the incorrect model of locality built into a particular interpretation of field theory.
 
Careful studies of entanglement and decoherence in virtual field fluctuations confirm the need to account for  causal consistency to avoid apparent paradoxes with nonrelativistic quantum thought-experiments\cite{Danielson2021,Belenchia2019}. 
The decomposition into modes that are used to construct a vacuum state--- the differentiation between radiation and vacuum---  depends on the choice of  Cauchy surfaces used to describe the  system.
Apparent paradoxes are resolved when field states are measured on Cauchy surfaces that correspond to correlated measurements.

Similar causal coherence of primordial virtual fluctuations  has recently been used  in a model that explains some observed anomalies of cosmological anisotropy at large angular separation \cite{Hogan_2022,Hogan2023}.
In that context,  directional hemispherical coherence leads to a causal ``shadow'' in primordial virtual correlation, which is observed as a symmetry of temperature correlations at large angles.

\subsubsection{Why now?}


 
Typically, field-based models of cosmic acceleration require introduction of new fields with new, arbitrary and very small dimensionless parameters, in some cases accompanied by an anthropic explanation\cite{Weinberg:1988cp}. 
In the gluonic-bubble scenario,  $\Lambda$  is not an independent parameter, but should have a precisely calculable value from Standard Model field fluctuations and standard semiclassical gravity.

In principle, this scenario roughly  accounts  for  the  well-known  puzzle sometimes nicknamed  the   ``why now'' coincidence--- the fact that the timescale associated with fundamental cosmic acceleration coincides  with the current age of the universe, which in turn presumably is determined by  astrophysical timescales, such as those determined by  stellar evolution.
The very long evolution timescale of  stars and other astrophysical systems in Planck units  originates mainly from   the cube of the nucleon mass\cite{Hogan:1999wh}:
\begin{equation}
T_{astro}/t_P \sim (m_P/m_{proton})^{3}.
\end{equation}
(Additional dimensionless factors that are numerically less significant, such as the electromagnetic  coupling and electron/nucleon  mass ratio, depend on the specific astrophysical system.)
The exponentially large dimensionless number $m_P/m_{proton}$, which expresses the weakness of gravity on a nuclear scale,  appeared mysterious to Planck, Eddington and Dirac, but now  has a  natural interpretation in the context of modern unified field theory,  because of the logarithmic running of the QCD coupling constant with energy scale\cite{Wilczek1999}.
In any case, because nucleon masses are  determined by the same scale that fixes  masses of pions and QCD vacuum fluctuation bubbles, the astrophysical timescale $T_{astro}$ naturally (roughly) coincides with the bubble model for $T_\Lambda$, since they originate from the same
 large  dimensionless number.

  \section{Conclusion}
  
   The  gluonic bubble  model demonstrates the classical gravitational coupling of a single pion-like oscillation with geometry. Our proposal for the cosmological constant is that fluctuations in the QCD vacuum have a similar relationship with gravity. Delocalized zero-point fluctuations of field vacua contribute nothing to the mean density that couples to gravity, but   locally coherent fluctuations have a net repulsive effect that mimics a uniform cosmological constant.  This effect occurs for the strong interactions in particular because of  gluonic tension, represented in our toy model by highly tensile gluonic gas. Our model shows  how this works geometrically  in classical systems, and why a  nonlocally coherent 4D structure is needed to obtain a net repulsive gravitational effect.

Although the  bubble model adopted here is a simplified  idealization of  real QCD vacuum states, the essential elements that create the cosmological
 constant of the magnitude estimated here---   nonlocal directional causal coherence of vacuum states, and a  tension from the strong nonabelian self-interactions of gluon fields--- must also  appear in the states of the physical QCD vacuum.
In this scenario, the absolute value of the physical cosmological constant can in principle be calculated exactly from a nonlinear computation of  spacelike correlations of 4D mass-energy flows in  the virtual QCD vacuum. 
 Such a calculation would allow more precise tests than the approximate agreement obtained here with a highly  idealized picture.



\bibliography{CCbib}

\section{Appendix}

\subsection{Gravity of a gluonic string model}
\subsubsection{Nonrotating string model}

We now consider the gravitational effect if the gluonic material does not uniformly fill a causal diamond as it does in the bubble model. 
Consider an extended  particle of total mass $M$ represented by two ``quarks'' of mass $m<<M$ connected by a one dimensional  string with  mass per length $\mu=\mu_0 M^2 (c/\hbar)$, where  $\mu_0$ is a dimensionless parameter that characterizes the gluonic tension. The system has zero angular momentum, so the quarks travel on radial trajectories connected by the straight string. It starts with  zero length, stretches to length $L_M= \hbar/\mu_0 Mc$, then re-contracts under the string tension.

For a classical model of strongly interacting gluonic quantum field excitations like pions, we can take the parameter $\mu_0$ to be less than but  of the order unity. Values $\mu_0<<1$ correspond to longer strings, which break apart into many shorter strings as new quark pairs are created.  Values $\mu_0>1$ describe systems smaller than the de Broglie wavelength, so they do not correspond to physical quantum states.

As in the bubble model,  very light quarks of mass $m<<M$, with  a negligible fraction of the total mass,  start with very high gamma-factor $\gamma= M/2m$ relative to the center, so the trajectories are nearly null.  
Their kinetic energy is converted to string energy as they travel. Eventually $\gamma\approx 1$, and for a  time short compared to $L_M/c$,  they enter a subrelativistic regime where the string tension turns them around.    For pion-like systems with $m<<M$, the short subrelativistic regime is a small fraction of the whole trajectory, so  the main effects do not depend strongly on the value of $m$.

 \begin{figure}[hbt]
  \centering
  \includegraphics[width=.4\textwidth]{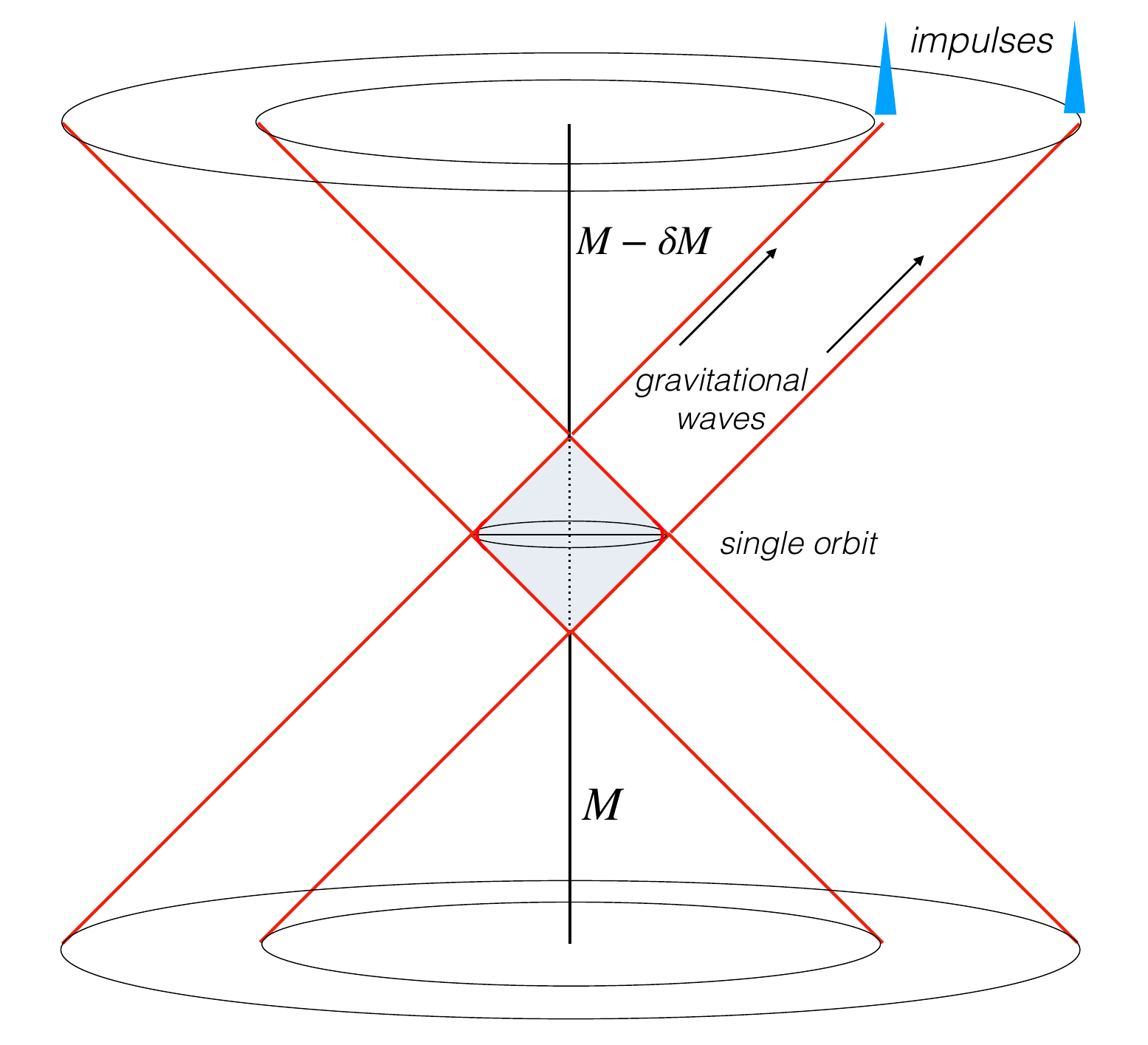}
  \caption{Causal diagram of gravitational radiation from a single gluonic string orbit. For $M>>m$, displacement is concentrated near anisotropic, spherical null shocks. 
  \label{causalwaves}}
\end{figure}

   \subsubsection{Gravitational waves}
   
  As shown in more detail in the linear solution below,  a gluonic string produces gravitational waves (Fig. \ref{causalwaves}).
   A simple estimate of the radiation rate from dimensional arguments shows how  the rate of energy loss scales with $M$. 
   In one orbit, a fully relativistic string with deficit angle $\delta \theta$  radiates a  fraction
   \begin{equation}
  \delta\theta\sim \delta M/M \sim \mu G/c^2  \sim \mu\hbar/c m_P^2
   \end{equation}
    of its energy as gravitational radiation.
    For a gluonic string of length $L_M = \hbar/\mu_0 Mc$ with $m<<M$,  the  decay rate by gravitational radiation in Planck units is about
   \begin{equation}\label{planckrate}
      t_P/ \tau_d\sim  \mu_0^{2} (M/m_p)^3.
   \end{equation}
   The string model displays roughly the same energy flow between QCD fluctuations and the gravitational vacuum as the bubble model, but it does not produce the same mean classical repulsion as gravity from bubble-like  fluctuations.
  
  \subsubsection{Linear string solution for $m\ge 0$}

   \subsubsection*{i. Equations of motion}
   
   For all computations in this section we will work in units where $c=1,G=1$ and restore units when needed. The energy-momentum tensor
   \begin{equation}
       T^{ab}=T^{ab}_s+T^{ab}_p
   \end{equation}
   consists of two parts: that of particles (representing quarks),
   \begin{equation}
       T^{ab}_p= m\delta(x)\delta(y)(\delta(z-z_1(t))v^a_{1}u^b_{1}+1\leftrightarrow 2),
   \end{equation}
 whose four-velocity is
  \begin{equation}
      u_a=\gamma v_a = \gamma_{1,2}(1,0,0,\dot{z}_{1,2}),
  \end{equation}
   and that of a string (representing gluons),
   \begin{equation}
    T^{ab}_s=\mu\delta(x)\delta(y)\mathcal{T}^{ab}\Theta(z-z_2(t))\Theta(z_1(t)-z),
   \end{equation}
   where $\mu$ is the constant mass per unit length of the string, $\mathcal{T}^{ab}\equiv t^at^b-z^az^b$,  and $\Theta(x)$ is the heaviside step function which is $1$ for $x>0$ and vanishes elsewhere.

   The equations of motion  are similar for the two models, the string and  the bubble.
   They are not identical: in the bubble, the quark surface mass density decreases with radius, so the inward acceleration at the edge for  constant $p$ increases with radius.
   The difference between solutions is small for $m<<M$, since the turnaround happens within a small fractional change in radius.

   In the string model, conservation of total stress-energy $\nabla_a T^{ab}=0$ gives an equation of motion for the positions of the quarks $z_{1,2}(t)$,
   \begin{equation}\label{stringquarkeom}
       \frac{d}{dt}(\gamma_{1,2} \Dot{z}_{1,2})=\mp \frac{\mu}{m} \equiv \mp \alpha.
   \end{equation}
   The solution for a uniformly accelerated relativistic particle with constant (proper) acceleration $\alpha$, initial 3-velocity $v_0$, and initial position $z_0=0$ is given by
   \begin{equation}\label{stringquarktraj}
       z_{1,2}(t)=\pm \frac{1}{\alpha}[ \gamma_0-(1+(\gamma_0v_0-\alpha t)^2)^{1/2}].
   \end{equation}
   The acceleration parameter is given by $\alpha=\mu/m$.
   Note that the turnaround time (or equivalently, the duration of time for which the particle is non-relativistic) is approximately $\delta t\sim 1/\alpha \sim L/\gamma_0$. 
   

   In the string model, we will see later that the peaks in the ``radiative'' part of the curvature occur at the retarded time associated with the turnaround of each particle. The period of the trajectory is $T=2\gamma_0v_0/\alpha$. Therefore the fraction of time during which the trajectory is non-ultra-relativistic is $\delta t/T\sim 1/\gamma_0 v_0\sim 1/\gamma_0<<1$. However, we will see later that the relative acceleration experienced by nearby test bodies scales like $\dot{v}\sim \alpha^2$, so that $\delta v\sim \dot{v}\delta t \sim \alpha$. 
   
   
   \subsubsection*{ii. Gravitational effect}
   To determine the gravitational effect of such a system, we begin with the linearized Einstein equations in the Lorenz gauge $\nabla^a \bar{h}_{ab}=0$, where $\bar{h}_{ab}$ is the trace reversed metric perturbation. 
   \begin{equation}\label{linearizedeinstein}
       \nabla_c\nabla^c \Bar{h}_{ab} = - 16\pi T_{ab}
   \end{equation}
   The solution to this equation is found by integrating over the intersection of the source world sheet with the past light cone. 
   \begin{equation}\label{metricintegral}
       \bar{h}_{ab} = 4\int_{\Lambda} \frac{T_{ab}(t',\vec{x}')}{|\vec{x}-\vec{x}'|} d^3 x'
   \end{equation}
   where $\Lambda$ denotes the past light cone of the event $(t,\vec{x})$. Since the delta function depends both explicitly on $\vec{x}'$ and implicitly through $t_{\rm ret}$, we need to use the Jacobian of the coordinate transformation to evaluate the integrals of the delta functions. 
   
   
   The metric perturbation associated with the low mass particles is
   \begin{equation}\label{particlemetric}
       \bar{h}^p_{ab} = \frac{4m\gamma_1v^{1}_av^{1}_b}{\alpha_1|\vec{x}-\vec{X}_1(t_{\rm ret})|} + \frac{4m\gamma_2v^{2}_av^{2}_b}{\alpha_2|\vec{x}-\vec{X}_2(t_{\rm ret})|}
   \end{equation}
   where
   \begin{equation}
       \alpha_{1,2} = 1 - \hat{n}_{1,2} \cdotp \frac{d\vec{X}_{1,2}}{dt}(t_{\rm ret})
   \end{equation}
   \begin{equation}
       \hat{n}_{1,2} = \frac{\vec{x}-\vec{X}_{1,2}(t_{\rm ret})}{|\vec{x}-\vec{X}_{1,2}(t_{\rm ret})|}
   \end{equation}
   From this point forward, evaluation at retarded time will be understood unless explicitly stated otherwise. The metric perturbation associated with the string is
   \begin{equation}\label{stringmetric}
       \bar{h}^s_{ab} = 4\mu {\rm ln}\left(\frac{z-z_2+\sqrt{s^2+(z-z_2)^2}}{z-z_1+\sqrt{s^2+(z-z_1)^2}}\right)\mathcal{T}_{ab}
   \end{equation}
   where $s^2=x^2+y^2$ denotes the transverse distance from the string. We note here that while in the case of the infinite string the spacetime is flat with an angular deficit, the dynamic string does in fact produce curvature. We are most interested in the leading order in $1/r$ behavior of the metric perturbation. In this limit, $\hat{n}\rightarrow \hat{x}$ and $t_{\rm ret} \approx t-r+{\rm cos}\theta z_{1,2}(t_{\rm ret})$.
   \begin{equation}\label{farfieldstringmetric}
       \bar{h}^s_{ab} = \frac{4\mu}{r}(z_1-z_2)\mathcal{T}_{ab} + \mathcal{O}\left(\frac{1}{r^2}\right)
   \end{equation}
   \begin{equation}\label{farfieldparticlemetric}
       \bar{h}^p_{ab} =\frac{4m}{r}\left(\frac{\gamma_1 v^{1}_av^{1}_b}{1-{\rm cos}\theta \dot{z}_1(t_{\rm ret})}+1\leftrightarrow 2\right) + \mathcal{O}\left(\frac{1}{r^2}\right)
   \end{equation}
   In the far field limit, the retarded time can be solved explicitly as a function of $u=t-r$ and $\theta$. 
   \begin{multline}\label{rettime}
       t^{1,2}_{\rm ret} = \frac{1}{\alpha {\rm sin}^2\theta}\left( \vphantom{\sqrt{u^2}} u \alpha \pm \gamma_0 {\rm cos}\theta \right.  \\
       \left.\mp \left|{\rm cos}\theta\right|\sqrt{({\rm sin}^2\theta+(u\alpha \pm \gamma_0{\rm cos}\theta)^2)}\right)
 \end{multline}
   
\begin{figure}
       \centering
       \includegraphics[width=\linewidth]{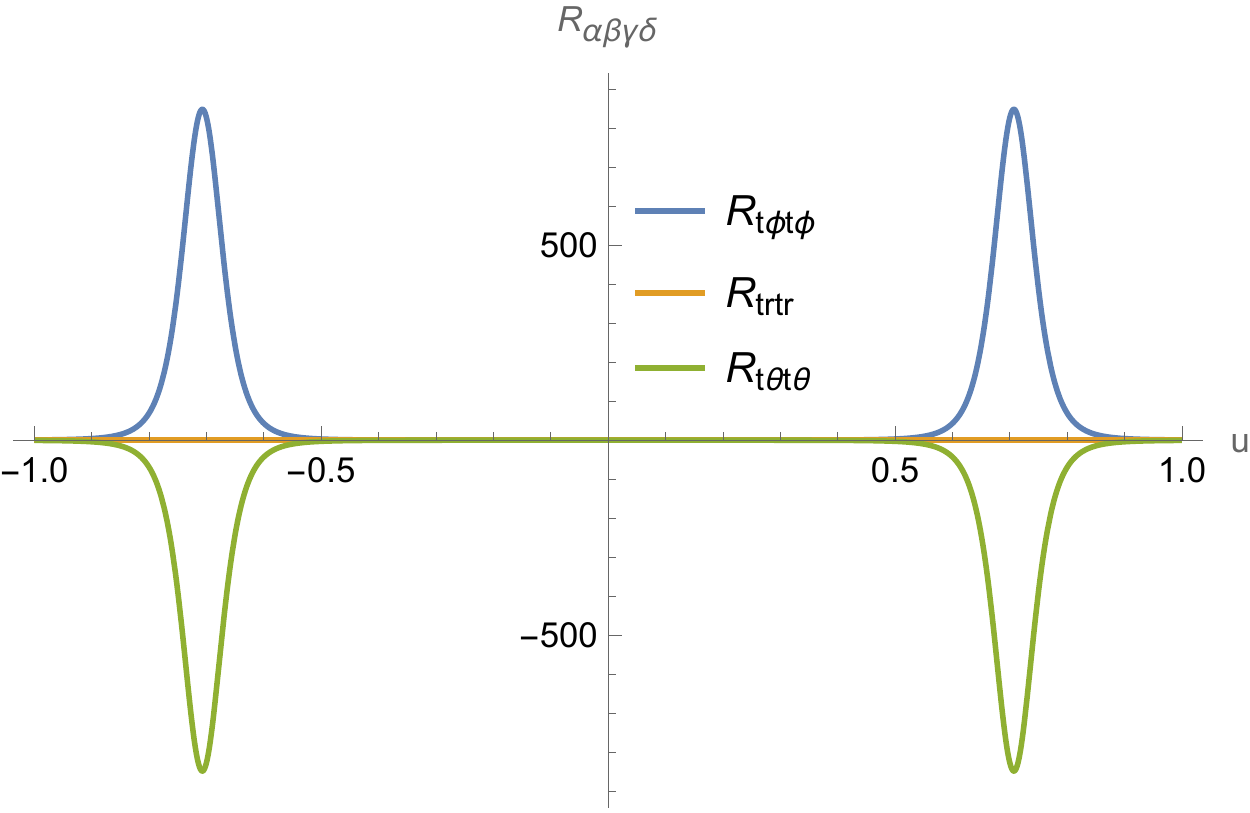}
       \caption{Components of the Riemann curvature tensor from eq. (\ref{geoddeviation}) plotted as a function of retarded time $u=t-r$ for $\gamma_0=10,\alpha=10,\theta=\pi/4$. The curvature spikes at the retarded times associated with the turnaround of each quark. Due to the highly relativistic motion, the turnaround is no longer simultaneous as seen by an observer away from the equator ($\theta=\pi/2$). }
       \label{fig:stringcurvature}
   \end{figure}

   The linearized Riemann tensor is given in terms of the metric perturbation by
   \begin{equation}
       R_{abcd} = 2\nabla_{[a}\nabla_{|[d}h_{c]|b]}
   \end{equation}
   The geodesic deviation equation tells us how nearby test bodies move relative to one another for small separations. 
   \begin{equation}\label{geoddeviation}
       t^c\nabla_c(t^d\nabla_d D^a) = - R^a_{\;cbd}t^ct^dD^b
   \end{equation}
   Since we are primarily interested in the far field behavior of the curvature, we need only consider derivatives which act on the time dependent terms in eq. (\ref{farfieldstringmetric}) and (\ref{farfieldparticlemetric}). One can show that $\partial t_{\rm ret}/\partial t = 1/\alpha_{1,2}$ and $\nabla t_{\rm ret} = - \hat{n}/\alpha_{1,2}$. In other words, $\nabla_{a}f(t_{\rm ret})=-K_{a} \dot{f}(t_{\rm ret})/\alpha_{1,2}$, where $K_{a}=-\nabla_{a}u = t_{a}+r_{a}$ is a radially out-going null vector. In this limit, the linearized Riemann tensor simplifies to
   \begin{equation} \label{farfieldcurv}
       R_{abcd} \approx 2 K_{[a}K_{|[d}\ddot{h}_{c]|b]}
   \end{equation}
   The relevant components of the Riemann tensor for eq. (\ref{geoddeviation}) are plotted in Fig. \ref{fig:stringcurvature} as a function of the retarded time coordinate $u=t-r$. Note the sharp spikes in curvature, which occur at the turnaround points of the quarks as viewed by a stationary observer.
   
\begin{figure}[h]
     \centering
    \includegraphics[width=\linewidth]{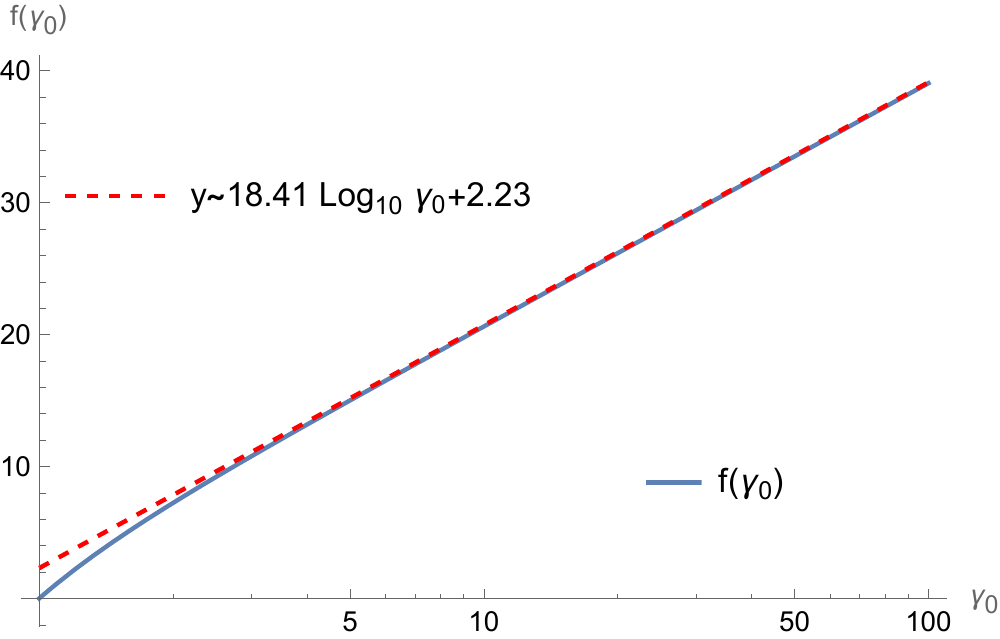}
     \caption{Plot of $f(\gamma_0)$ in eq. (\ref{gravwavepowerscaling}) on a log-linear scale, demonstrating an asymptotically logarithmic scaling of the power. }
     \label{fig:powerscaling}
 \end{figure}

   The effective stress-energy tensor for the gravitational waves is given by
   \begin{multline}\label{gravwavestress}
       T^{\rm GW}_{ab} = \frac{1}{32\pi}\left\langle \Bar{h}_{cd},_a\Bar{h}^{cd},_b-\frac{1}{2}\Bar{h},_a\bar{h},_b \right.  \\
        - \left. \vphantom{\frac{1}{2}}\Bar{h}^{cd },_{d}\Bar{h}_{c b},_{a}-\Bar{h}^{cd},_{d}\Bar{h}_{c a},_{b} \right\rangle
   \end{multline}
   where $\langle ...\rangle$ denotes an average over several wavelengths of the radiation. Although this object is not gauge invariant (and therefore its physical interpretation is unclear), the total integrated flux of energy to null infinity as defined by
    \begin{equation}\label{gravwavepower}
       P = -\lim_{r\rightarrow \infty} \int T_{0a}dS^a
    \end{equation}
    is gauge invariant and is therefore a physically meaningful quantity. 
   It turns out that the dynamics of the string do not contribute directly to the stress energy of the gravitational waves in this setup, although the string does indirectly contribute by determining the dynamics of the quarks. Restoring units, the total integrated power in the string model scales like
\begin{equation}\label{gravwavepowerscaling}
       P= f(\gamma_0) \alpha^2 m^2 G c^{-1}
 \end{equation}
  where $f(\gamma_0)$ is plotted in Fig. \ref{fig:powerscaling} on a log-linear scale. We see that the asymptotic scaling of $P$ is logarithmic in $\gamma_0$, hence confirming the previous assumption that the power does not strongly depend on the mass of the quarks in the $m<<M$ limit. For a pion mass of $135$ MeV and quark masses of about $3-5$ MeV, we get $\gamma_0 = M/2m \sim 10^1$ and $f(\gamma_0)\sim 10^1$.  In agreement with the estimate given by eq. (\ref{planckrate}),  the timescale of decay is
 \begin{equation}
     \tau_d^{-1} = \frac{P}{M}\sim 10^1 \mu_0^2 (M/m_p)^3 \tau_p^{-1} 
 \end{equation}

 \subsection{Strong energy condition, quantum trace anomaly, and dimensional dependence}

We now explore how the particular nature of the QCD interactions  can  allow for a repulsive gravitational effect. Consider a congruence of timelike geodesics described by a vector field $u_a$. The expansion, shear, and twist of the congruence are defined by
\begin{align}\label{expansion}
    \theta &= q_{ab}\nabla^au^b \\
    \sigma_{ab} &= q_a^{\,c}q_b^{d\,}\nabla_{(c}u_{d)}-\frac{1}{3}\theta q_{ab} \\
    \omega_{ab} &= q_a^{\,c}q_b^{\,d}\nabla_{[c}u_{d]}
\end{align}
where $q_{ab}$ denotes the spatial part of the metric that is orthogonal to $u_a$, and $(a,b),[a,b]$ denote symmeterization and antisymmeterization of indices, respectively. Raychaudhuri's equation tells us how the expansion evolves with time. 
\begin{equation}\label{raych}
    \frac{d\theta}{d\tau} = -\frac{1}{3}\theta^2-\sigma_{ab}\sigma^{ab}+\omega_{ab}\omega^{ab}-R_{ab}u^au_b
\end{equation}
Here $R_{ab}$ denotes the Ricci tensor. As is typically done, we will ignore the twist. For a congruence which initially has zero expansion and shear, we have (after using the Einstein field equations to replace $R_{ab}$ with $T_{ab}$)
\begin{equation}\label{rateofexp}
    \left.\frac{d\theta}{d\tau}\right|_{t=0} = -8\pi\left(T_{ab}-\frac{1}{2}g_{ab}T\right)u^au^b
\end{equation}
where $T=g^{ab}T_{ab}$, which evaluates to $T=-T_{00}+T_{11}+T_{22}+T_{33}$ in global inertial coordinates. For a perfect fluid in 3 spatial dimensions (in units where $c=1$), 
\begin{equation}
    T_{ab}={\rm diag}(\rho,p,p,p),
\end{equation}
 so eq. (\ref{rateofexp}) becomes
 \begin{equation}
     \left.\frac{d\theta}{d\tau}\right|_{t=0} = -4\pi(\rho+3p)
 \end{equation}
 Therefore if the perfect fluid obeys an equation of state $p<-\rho/3$ (such that the strong energy condition $R_{ab}u^au^b>0$ is violated), the expansion of the congruence will initially be positive, leading to a repulsive gravitational effect. As the expansion and shear grow, the RHS of eq. (\ref{raych}) will eventually reach zero, but the sign of the rate of change of expansion can never become negative.

 Now consider another fundamental constraint:  a perfect fluid composed purely of a classical non-abelian gauge field (e.g.,  massless noninteracting point particles, such as photons), the equation of state is $p=\rho/3$, which does not violate the strong energy condition. The form of the field strength tensor in classical Yang Mills theory is such that the trace of the stress-energy tensor always vanishes, similar to classical Maxwell theory. However, quantum interactions between the gluons and quarks and the self-interaction of gluons lead to what is known as a trace anomaly, i.e. a non-vanishing of the trace of the quantum stress energy tensor. 
 
 In order for the strong energy condition to be violated by quantum interactions, Eq. (\ref{expansion}) tells us that the equation of state must satisfy $\rho+T/2<0$, or $T<-2\rho$. 
  For a perfect fluid, we have $T=-\rho+3p$.  This puts a constraint on the magnitude of the trace anomaly if QCD vacuum fluctuations are to serve as the source of accelerating cosmic expansion.  It has been hypothesized \cite{Schutzhold:2002pr} that this trace anomaly can produce the violation of the strong energy condition needed to produce a repulsive gravitational effect.

 The vanishing of the trace of the stress-energy tensor is synonymous with conformal invariance of the action of a field theory, i.e. $S=\int d^4x\sqrt{-g} \mathcal{L}$ is invariant under $g_{\mu\nu}\rightarrow \Omega(x^{\alpha})g_{\mu\nu}$. In ref. \cite{Schutzhold:2002pr} it is shown that this conformal invariance is preserved under the standard QFT renormalization procedures for \textit{free} fields in flat space but \textit{not} for self-interacting fields in flat space or free fields in curved spacetimes. 
 It is estimated there that the self-interacting gluonic field gives rise to a negative energy density and pressure proportional to the renormalization group $\beta$ function, with a corresponding cosmological constant that is many orders of magnitude too large. 
 
According to our interpretation, where the gravitational effect of virtual fluctuations is evaluated using a causally coherent bubble model, the magnitude   is about right.  This result is consistent with the idea that   active gravity entangles long wavelength field modes with causal structure, which is not accounted for in standard renormalization procedures\cite{HollandsWald2004,Stamp_2015}.
 
 Although the sharp boundary and spherical symmetry of the bubble model are artificial, we conjecture that the physical gravitational effect is indeed microscopic,  localized at the Fermi scale. Unlike the model sketched in ref. \cite{Schutzhold:2002pr}, the cosmic acceleration is a true cosmological constant, with a value determined entirely by the properties of the stable QCD ground state vacuum. These arguments suggest that at very high temperatures in the early universe, when QCD had significantly weaker interactions, the value was much smaller, but there are no observable consequences of this variation.

 Ref. \cite{Schutzhold:2002pr} also shows that  free fields in curved space gives a positive energy density with a cosmological constant that is many orders of magnitude too small, which  supports our conjecture that fluctuations of the other Standard Model fields do not contribute
 significantly to the gravitational energy of the vacuum. The effects of the anomaly are nonperturbative, and are exponentially suppressed for coupling constants that are not of order unity.
 

 Finally, we address why the bubble model of the pion behaves differently from the string model for producing a repulsive gravity. Although the string model of the pion is commonly used as a toy model in other contexts, it does not work as a gravitational model  because of the dimensionality of the matter distribution. In order to violate the strong energy condition in $D$ spatial dimensions, the fluid must satisfy 
 \begin{equation}
T_{00}+\frac{1}{2}T=\rho+\frac{1}{2}(-\rho+Dp)<0 \rightarrow \rho+Dp<0
 \end{equation}
 Therefore, a pressure $p=-\rho$ in $D=1$ spatial dimensions (i.e. a straight string) cannot violate the strong energy condition: more than one dimension of tension is needed to account for cosmic acceleration.
 

\end{document}